# Atomic Structure of Polar Surface in SrTiO$_3$

Adeel Y. Abid[1,2#], Ning Li[1,2#], Ayaz Arif[3] and Peng Gao[1,2,4]*

[1]*International Center for Quantum Materials, School of Physics, Peking University, Beijing100871, China*

[2]*Electron microscopy laboratory, School of Physics, Peking University, Beijing 100871, China*

[3]*Department of Physics, University of Azad Jammu & Kashmir, Muzaffarabad 13100, Pakistan*

[4]*Collaborative Innovation Centre of Quantum Matter, Beijing 100871, China*

[#]*These authors contributed equally to this work.*

* *Author to whom the correspondence should be addressed to: p-gao@pku.edu.cn*

**Abstract**

Rearrangement of atoms due to broken translational symmetries at the surface of SrTiO$_3$ is scarcely debatable in the present day scenario. Actual concern demands to unveil the true structure and precise mechanism responsible for atomic reconstructions at a unit cell level. Here, we reveal the atomic structure of SrTiO$_3$ surface by using annular bright field imaging in spherical aberration corrected scanning transmission electron microscope. We measure structural parameters such as lattice constant, bond length, atomic displacement and subsequently we find a polarization for the top five unit cells is ~10–20 µC·cm$^{-2}$. Both Sr-O and Ti-O planes reconstruct thus contribute to net dipole moment that is towards surface. Elongation of the out of plane lattice for skin layers is also witnessed. The electron energy loss spectroscopy experiment illustrates the presence of oxygen vacancies at the surface which seems to be largely responsible for the stability of surface polarization. Our study provides a sound foundation into mechanistic understanding for structure and properties of surfaces for complex oxides. These outcomes can be considered as a pragmatic boost particularly for examining other insulator-like materials as well as in bridging gap for hitherto fewer experimental investigation along with various theoretical studies.

**Introduction**

Ferroelectricity in ABO$_3$ type perovskite oxides requires strong ionic and non-centrosymmetric phase which arises when the material is cooled below T$_c$. Exception in this connection is strontium titanate (SrTiO$_3$, STO) having no bulk ferroelectric phase at any finite temperature. However, ferroelectricity can be triggered by applying mechanical stress [1-4], by inducing electric field [5-8] or by creating defects such as Sr vacancies, antisite defects and oxygen vacancies [9-11]. Studies related to afore-mentioned phenomena are numerous, for instance, Weaver observed the existence of spontaneous polarization and hysteresis in STO below 50 K [5]. A decade later Burke *et al.* found the transition to the ferroelectric phase by applying uniaxial stress along more than one crystallographic orientations at 4.2 K [2]. Fluery and Warlock observed first order Ramen scattering by inducing electric field [6]. Besides, flexoelectric effect induced polarization was also monitored in STO single crystal [12]. Most of the studies have witnessed ferroelectric response at low temperatures. According to a report room temperature surface piezoelectricity in SrTiO$_3$ was also monitored by utilizing piezoresponse force microscopy [13]. Recently, room temperature ferroelectricity has also been reported for STO films grown on a slightly mismatched lattice DyScO$_3$ [14], and later at an even higher temperature (T=400 K) for STO films grown on silicon [15]. Even if the substrate strain is absent, the room temperature ferroelectricity was also observed in the ultrathin STO film [16].

Besides, surface ferroelectricity for a few top layers of strontium titanate has also been predicted [17-21]. Using low-energy electron diffraction (LEED), Bickel *et al.* found that the surface layer exhibits significant puckering at 120 K (oxygen ions extending away from the surface) [17]. Later theoretical work by Ravikumar *et al.* suggests that the surface can be ferroelectric if it is SrO-terminated but not if TiO$_2$-terminated [18]. Padilla *et al.* through first principle calculations of energy for [001] surface of STO provided some indication of weak surface ferroelectricity [19]. Surface X-ray diffraction investigation of strontium titanate surface carried out by Herger *et al*. gave an evidence about surface polarity by mapping atomic displacement down to three unit cells [20]. However, even after several theoretical and experimental studies, the atomic structure of surface is not so clear and whether or not the surface is polarized is still under debate mainly due to highly intertwined and complex structure at surfaces and most properties are directly linked to the structure which can easily be altered by small localized distortions at an atomic level [22]. To investigate these precise distortions and local structures, limited spatial resolution associated with bulk characterization techniques, to name a few, x-ray, neutron diffraction remained a chronic problem over the decades. Although the scanning tunneling microscopy (STM) and atomic force microscopy (AFM) are being considered as a powerful tools to observe surface reconstruction [23-25], but they provide only a top view and cannot perceive the subsurface layers consequently formation as well as the mechanism of skin reconstruction along with phenomena like polarization, strain and octahedral tilt, hence, left unexplored to a major extent [26-28]. In contrast, the recent advancements of annular bright field (ABF) imaging in an aberration corrected scanning transmission electron microscope (STEM) have made it possible to simultaneously determine both heavier cations and lighter oxygen atoms in oxides such as $TiO_2$ and $Al_2O_3$, allowing us to precisely measure the structural parameters at surface and subsurface of functional materials regardless of their poor electrical conductivities [29-31]. Thus, for the STO surface, by using this technique we are able to answer these mentioned-ahead questions that how large the polarization is, how much the thickness of the polarized layer is, and what is the possible contributions of SrO and $TiO_2$ layers to the total polarization.

In this paper, we use ABF in aberration corrected STEM to study the atomic structure of STO skin by mapping the lattice constant, bond length, atomic shift and polarization. Our result shows for a few top surface layers (~ five unit cells) the displacements of the oxygen and Ti relative to the center of the Sr sublattice occur at this surface layer, which generate a polarization as large as 10 to 20 $\mu C \cdot cm^{-2}$. Besides, the out of plane lattice constant c at surface has an increase of almost 3–4 % signifying a transition from cubic to tetragonal phase accompanied with emergence of polarization. The electron energy loss spectroscopy measurements identify the presence of oxygen vacancies that may account for the stable polarization. Our results reveal the atomic structure of surface region by confirming the presence of polarization, phase transition, and oxygen vacancies, providing valuable insights into the understanding of atomic structure and properties for complex oxides and shed light on design of devices via surface engineering.

**Results and Discussion**

Fig. 1(a) illustrates an ABF image of (100) $SrTiO_3$ thin film with a viewing direction of [010], from which both cation columns (Sr and TiO) along with lighter oxygen columns are visible. Therefore, this viewing direction is very suitable to capture any possible phase transition and subtle structure distortion. The topmost layer at left side is SrO labeled as #1 layer. Fig. 1(b) and 1(c) entail the magnified portion of three adjacent unit cells as marked by two rectangles in Fig. 1(a) for surface and bulk respectively and unambiguously displaying a difference in the context of atomic distortion which is almost rare for bulk but clearly distinguishable in case of surface as indicated by white outlines. The schematics of ideal cubic centrosymmetric unit cell of bulk STO is shown in Fig. 2(a) whereas Fig. 2(b) is an illustration of a unit cell at surface having ferroelectric tetragonal transition along c axis with distorted four-fold equivalent symmetry. These observations indeed argue about some signatures of polarity thus encouraging us to carry out quantitative analysis for atomic distortions and polarization.

After accomplishment, Fig. 2(c) is the plot for out of plane averaged atomic displacement mapped separately at both Sr-O and Ti-O planes ranging from surface (left side) to the deep inside bulk (right side). Next, we obtain the polarization quantitatively by calculating dipole moment per unit cell and variation of polarization for both skin and bulk in STO thin film is displayed in Fig. 2(d) (see Supplemental Materials for detail). The average value of spontaneous polarization for top four unit cells is $P_s$ ~10–20 µC·cm$^{-2}$ (See Supplemental Materials Fig. S1 for rest of the details). Identical trend of surface polarization is witnessed for same under investigated STO thin film by demonstrating the quantitative measurements of another surface region having spontaneous polarization ~10–15 µC·cm$^{-2}$ for the top 4 to 5 unit cells at the surface (details and related graphical illustration for atomic displacement and polarization are incorporated in Supplemental Materials Fig. S3).

Fig. 3 confirms the significant out of plane lateral expansion of lattice parameter c for top surface layers. For example, average value of lattice c calculated for top three unit cells at both SrO and TiO$_2$ plane is ~ 405.4 pm and ~ 403.1 pm respectively compared with the averaged value of 390.5 pm (as reference) below the 4$^{th}$ unit cell. There is no significant structural change along in plane direction as the lattice a remains unchanged along this orientation, as shown in Supplementary Fig. S5. EELS measurements are also carried out to track the possible changes in electronic structure at surface. In Fig. 4(a) the circles indicate the particular regions at which EEL spectra are recorded, and Fig. 4(b) is the corresponding spectra. Notably, peak splitting in the Ti L-edges is more pronounced and remain constant for bulk but at/or near surface less peak splitting in the Ti L-edges occurred as seen by four orange lines in Fig. 4(b), indicating that the Ti valance is partially reduced from $Ti^{4+}$ to $Ti^{3+}$ at surface [32]. In Fig. 4(c), the calculated average valence of Ti at the surface is ~3.63. In addition, a few peaks of O-K edge between 530 eV and 550 eV are visible in the STO bulk while they become flat and no longer distinguishable at the surface, which is an indication of oxygen vacancies [32]. For such oxides the formation of oxygen vacancies at surface is natural [33] and this phenomenon can be understood by the fact of weaker bonding of oxygen at the surface and thus they can easily escape from the surface, leaving oxygen vacancies at surface.

The ferroelectricity in STO is largely believed to originate from the naturally existed nanometer-size polar clusters (NPCs) due to the local inhomogeneities [34,35] such as Sr vacancies and Ti-antisites [36]. Indeed, the calculations suggested that the formation energy of Sr vacancies can be as small as that of the oxygen vacancies [37,38]. The recent first-principles calculations also revealed that Ti-antisite induced polarization is as large as 55 $\mu C/cm^2$ in its residing unit cell and also causes other surrounding region to be coherently polarized [16]. Although, the NPCs commonly exist in STO bulk single crystals and films but such small isolated NPCs are buried in an insulating bulk matrix and thus the unscreened depolarization field destabilizes their polarization. Therefore, they usually not remain capable of generating macroscopic polarization. However, such a delicate balance can easily be disturbed by external parameters such as the electrical fields [6,7], substrate strain in thin films [14,15], and charges from the defects [9,11], leading to a transformation from paraelectric STO to ferroelectric phase. At the STO surface, due to weakly bonded surface atoms cationic point defects are readily formed to breed high density of NPCs. The oxygen vacancies and possible foreign adsorbates on the surface can therefore act as a compensation charges to screen the depolarizing field and to align the electric dipoles, consequently stabilizing the long-range correlation among NPCs to produce macroscopic polarization. Moreover, Chisholm *et al.* also concluded that compensation mechanism at the interface of $PbZr_{0.2}Ti_{0.8}O_3/SrTiO_3$ is largely derived by the charges generated by oxygen vacancies [39]. This scenario is largely similar with that in another previous report by Lee *et al*. that the defective ultrathin STO film exhibits spontaneous ferroelectricity [16] because the surface and interface both can provide electronic or ionic charges for screening and thus impart into stabilization of polarization.

The atomic structure of the polarized surface can be extracted from the ABF image which depicts that octahedral shift (ferroelectric phase) is dominated at the surface of STO. Bond length among the few top surface layers is shown in Supplemental Table 1 (see Supplemental Materials for detail). By comparing with bulk value ($d_o$=195.25 pm) we found compression for first two layers whereas second and third layer seems to be expanded and also consistent with Fig. 1 wherein an overall towards bulk shift of oxygen atoms in third layer can be observed. Previous theoretical studies [19,40-42] either for Sr-O or Ti-O termination agrees upon this argument but with exception of magnitude. However, REED investigations carried out by Hikita *et, al.* shown expansion for both top layers spacing and this discrepancy is probably due to a little influence of RHEED intensities on $d_{23}$ [43]. The average value of bond lengths down to three unit cells is ~201.9 pm with a relaxation of ~3.4 % along [001] direction. Knowing the fact that in bulk relative atomic displacements vanish but appear at surface due to lowered symmetry. Generally speaking, in perovskite oxide-cation displacement is directly proportional to the ferroelectric polarization with pronounced effect [44]. We found that Ti and Sr atoms in top four unit cells are displaced from their centrosymmetric place with an average displacement of ~12.9 pm and 9.2 pm respectively along with relative shift of oxygen atoms in their respective Sr-O and Ti-O layers for top four unit cells, which is ~9.1 pm in case of premier and for latter it is ~13.2 pm. Ultimately, by mainly utilizing the atomic shift/displacement data we found an average value of spontaneous polarization for top four unit cells is $P_s$ ~10–20 μC·cm$^{-2}$ (see Supplemental Materials for detail).

Beside quantitatively judging polarization from atomic displacement, we may also further verify its occurrence from lattice expansion ~3.5% for top four unit cells. In fact, the elongation of lattice c observed at different regions of the surface (see Fig. S4 Supplemental Materials for other surface region) is also consistent with previous observations for other perovskites [45,46]. According to Landau-Ginsburg-Devonshire (LGD) theory the equilibrium polarization and lattice c are co-related to each other [47], i.e., by using lattice parameters we may calculate polarization. This scenario entails with certainty that the phase transition from cubic to tetragonal at few nanometers thick STO surface is accompanied with generation of polarization. Moreover, this argument of lattice elongation possesses a deep link with our another vital finding of O vacancy presence at surface layers by viewing through the nexus of Vegard effect[48] according to which ion concentration variation induces stress or strain means there is an effect of stoichiometry on the lattice constant. Hence, it can be securely asserted that lattice expansion phenomenon is a signal to O vacancies presence which in turns act as a possible compensation charges to screen the depolarizing field and to align the electric dipoles thus contribute for polarization.

The atomic structure is schematically presented in Fig. 2(a) and 2(b) indicating the polarization is outward, i.e., polar vectors pointing to surface. The polarization direction is different than that predicted in a LEED investigation by the calculation for the ideally clean surface in the vacuum [17] because the orientation of skin polarization is determined by the practical surface conditions [46], e.g., dangling bonds, point defects, and adsorbates, which indeed are difficult to identify by our method and needs further investigation in future. Our current investigation of surface polarization is, of course, not for ideally clean surface instead the polarization reflects true information for practical materials as most of previous characterization of STO surface polarization such as PFM or AFM and electrical measurements [13,21,49,50] were also done in atmosphere, during which similar surface amorphization, absorption and contamination were also evitable. Indeed, previous studies reported that the surface chemical environment can even reversibly switch the polarization direction in ferroelectric thin films [51] proving the external adsorbate can be critical to control the polarization orientation.

Furthermore, due to the reconstruction at skin layers different physical and chemical properties arises which significantly influence and alter the response of ferroelectric devices particularly in ultra-thin films [52], for example, surface photochemistry and photocatalysis applications such as dye degradation and water splitting [53]. Already, silicon industry has considered a replacement of $SiO_2$ as a gate dielectric with alternative high 'k' materials such as $SrTiO_3$ and for successful accomplishment appropriate quality and structural compatibility of the interface between the two materials is required [54]. Hence, understanding the structural features and formation mechanisms of STO surfaces plays a noteworthy role in this endeavor.

In summary, we demonstrate the profound insight mechanism of spontaneous polarization at STO surface. The thickness of polarized surface layer is about five unit cells and the average polarization is ~10 to 20 $\mu C \cdot cm^{-2}$. The polar vectors orient towards surface hence constructing a positively poled surface. Both the planes either Ti-O or Sr-O reconstruct, however with distinct scale, and participate for net dipole moment. Accompanied with the generation of polarization at the surface, the structure transforms from original cubic into tetragonal with prominent expansion ~3.4% in lattice c. Through EELS spectra, we confirm the presence of O vacancies at the surface, which is in good agreement with the lattice expansion according to Vegard effect. These O vacancies are found largely responsible for stabilization of net polarization in a non-ferroelectric strontium titanate. Polarized surface of STO have potential applications in surface catalysis and future electronic devices. Our experimental methodology demonstrated in this work enabling picometre-scale measurement of structural parameters of both surface and bulk can be greatly helpful in comprehending surface structures and properties in the low dimensional functional materials particularly for other insulators, which may be not suitable for the STM study as well as explores novel horizons into engineering of ferroelectric surface for the surface chemistry applications.

Acknowledgement

P.G. is grateful for the use of Cs corrected TEM supported by Prof. Yuichi Ikuhara and technical support from his group members in University of Tokyo. The authors acknowledge the support from the National Basic Research Program of China (2016YFA0300804), National Natural Science Foundation of China (51672007, 51502007, 11327902), National Equipment Program of China (ZDYZ2015-1), and the National Program for Thousand Young Talents of China, and "2011 Program" Peking-Tsinghua-IOP Collaborative Innovation Center for Quantum Matter.

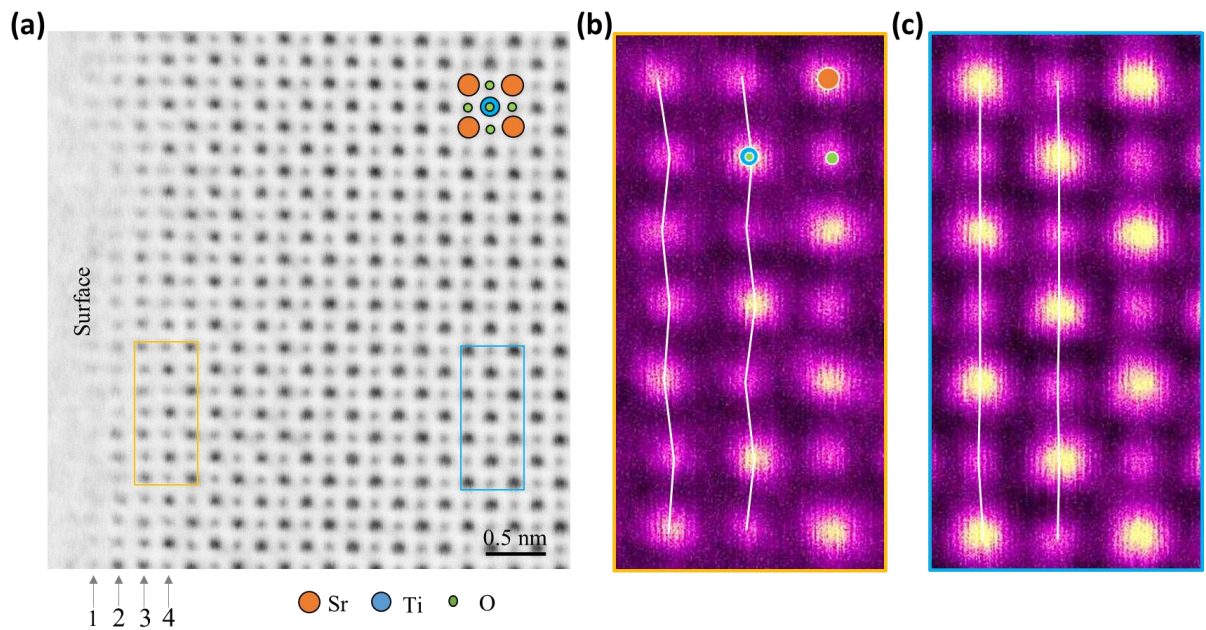

**Fig. 1** Atomic structure of SrTiO₃ surface. (a) An atomically resolved annular bright-field (ABF) image of SrTiO₃ (STO) surface. A schematic for this ABF image is depicted at up-right corner. The large-size orange circles, medium-size blue circles and small-size green circles show the atomic positions for Sr, Ti and O respectively. The atomic layers are labelled with the numbers and viewing direction is along [010]. The enlarged view of two groups of adjacent unit cells as marked by rectangular regions in "a" taken at (b) surface and (c) bulk of STO.

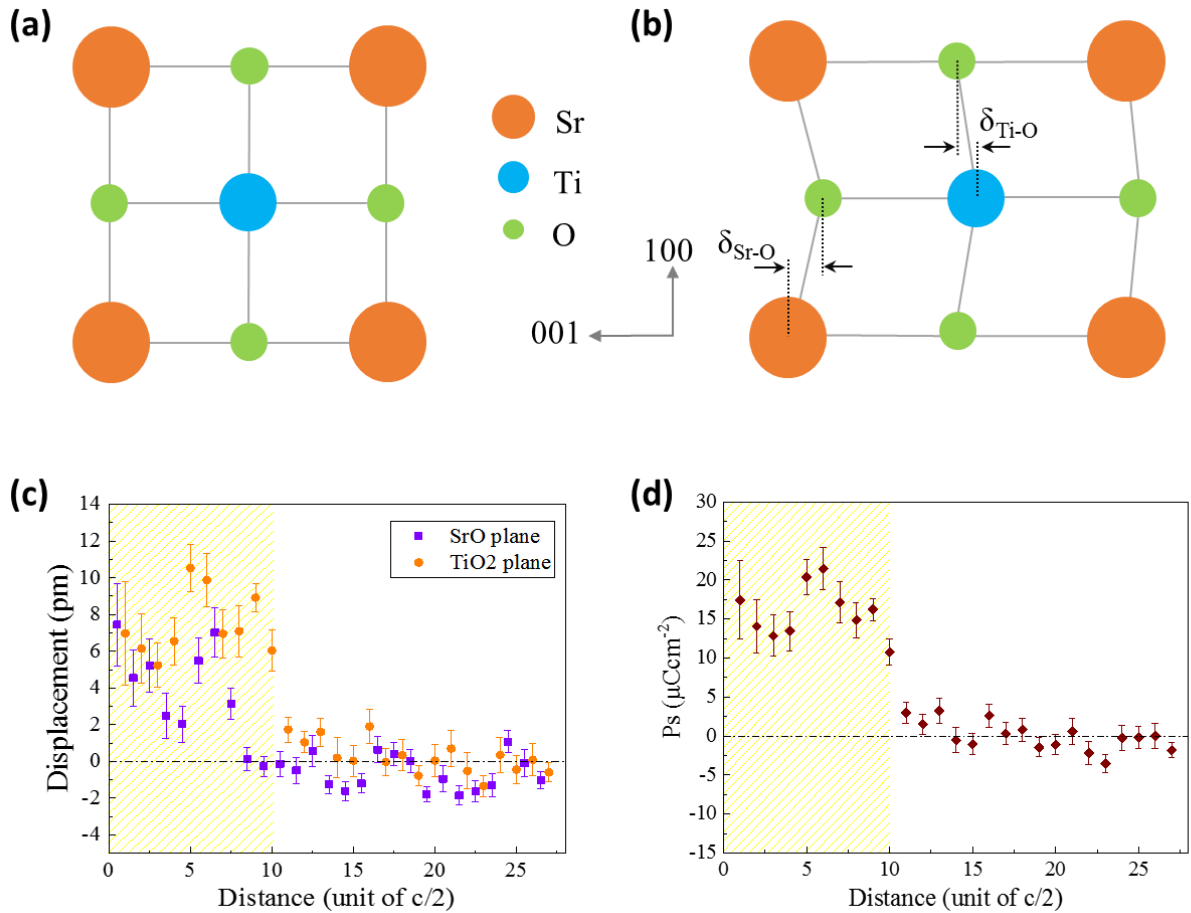

**Fig. 2** Illustration of schematics and quantitative measurements of structural parameters such as atomic displacement and polarization for STO thin film. The area of calculated region includes 16×13 number of unit cells. (a) The schematics of ideal cubic centrosymmetric unit cell of bulk STO monitored along the crystallographic b axis (Sr: orange, Ti: blue, O: green). (b) Proposed schematics of STO unit cell at surface having ferroelectric tetragonal phase viewed along [010] direction where $\delta_{Sr-O}$ and $\delta_{Ti-O}$ are the relative out of plane (along c axis) atomic shift of Sr, Ti and O atoms from centrosymmetric origin at their respective planes. (c) Plot of z-component ([001] direction) of atomic displacement at both SrO and $TiO_2$ planes along c axis. The purple squares denote SrO plane whereas orange circles show $TiO_2$ plane. The error bar is the s.d. (d) Plot of total spontaneous polarization obtained from the displacement data. The error bar is the s.d.

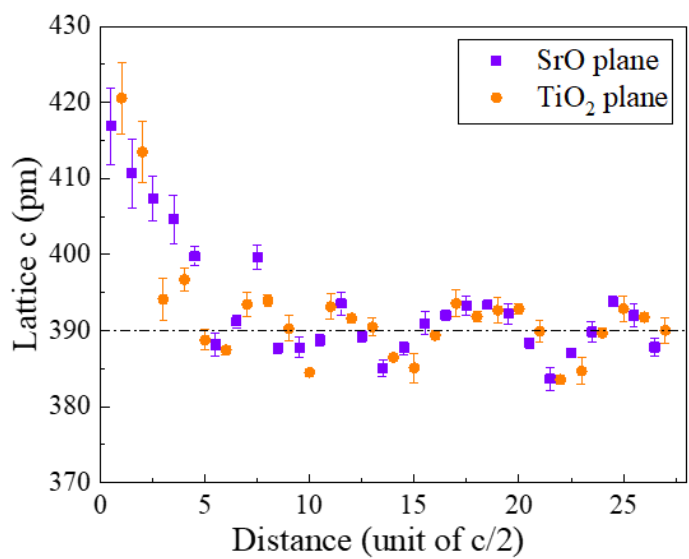

**Fig. 3** Trend of lattice elongation along crystallographic orientation c for under investigation STO thin film. Plot for mean of the lattice parameter in [001] direction. Twenty-seven data points where averaged for each plane to calculate lattice c. The error bar is the s.d. The purple squares denote SrO plane whereas orange circles show $TiO_2$ plane. Significant expansion of lattice c is being witnessed for skin as compared to bulk.

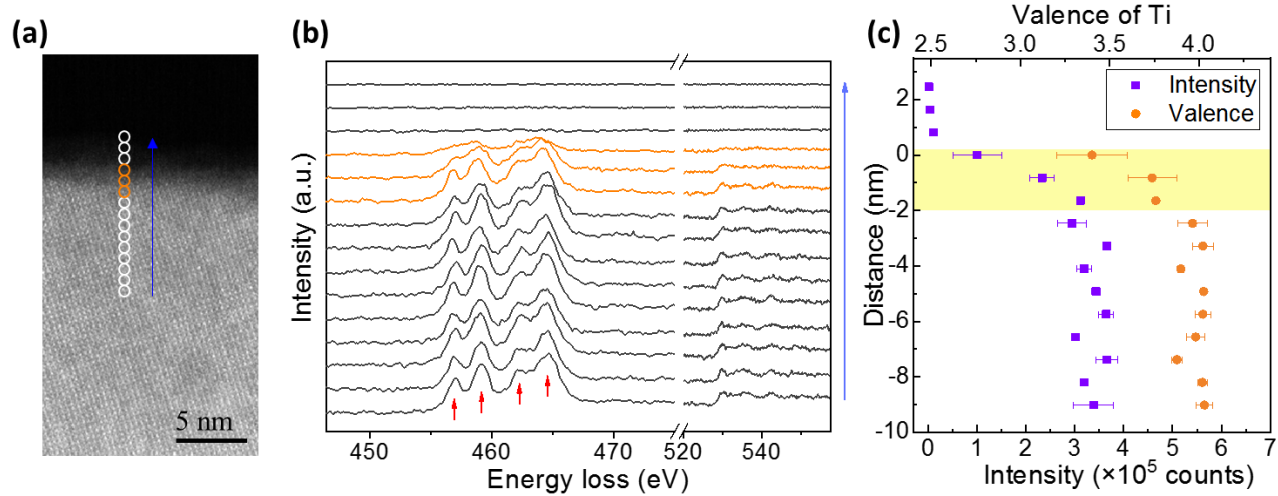

**Fig. 4** Electron energy loss spectroscopy (EELS) measurement. (a) A HAADF image. The circles highlight the locations to record EEL spectra. The blue arrow indicates the scanning direction. (b) The corresponding EEL spectra. The orange spectra from surface show less peak splitting, indicating the presence of oxygen vacancies at the surface. (c) Change of Ti valence and the intensity of Ti-L edges. The purple squares denote intensity calculated from the integral of Ti peaks whereas orange circles show Ti valence calculated from the EEL spectra. The error bar is the s.d.

**Atomic Structure of Polar Surface in SrTiO$_3$**

Adeel Y. Abid[1,2,#], Ning Li[1,2,#], Ayaz Arif[3] and Peng Gao[1,2,4]*

[1]International Center for Quantum Materials, School of Physics, Peking University, Beijing100871, China

[2]Electron microscopy laboratory, School of Physics, Peking University, Beijing 100871, China

[3]Department of Physics, University of Azad Jammu & Kashmir, Muzaffarabad 13100, Pakistan

[4]Collaborative Innovation Centre of Quantum Matter, Beijing 100871, China

[#]These authors contributed equally to this work.

* Author to whom the correspondence should be addressed to: p-gao@pku.edu.cn

**Supplemental Materials**

1. **TEM sample preparation**

   In order to prepare sample for ABF image acquisition we first used mechanical polishing and then argon ion milling. The ion-beam milling was carried out using PIPSTM (Model 691, Gatan Inc.) with the accelerating voltage of 3.5 kV until a hole is made. Low voltage milling was performed with accelerating voltage of 0.3 kV to remove the surface amorphous layer and to minimize damage.

2. **Image acquisition and Data analysis**

   HAADF and ABF images were recorded at 300 kV in a JEM ARM300CF (JEOL Ltd.). The convergence semi-angle for imaging is 24 mrad, the collection semi-angles snap is 12 to 24 mrad for the ABF imaging and 65 to 240 mrad for the HAADF imaging. The EELS experiments were carried out in a JEM ARM200CF (JEOL Ltd.) equipped with dual Enfinium camera (Gatan). The spectra were recorded at 200 kV. The electron beam was slightly spread to minimize the possible effects under electron beam illuminations, and the acquisition time is 0.5 s/pixel with energy dispersion 0.1 eV. The atomic positions in Fig. 1(a) were determined by simultaneously fitting with two-dimensional Gaussian peaks to a perovskite unit cell using a MatLab code [1] through which we obtained x, y coordinates of each atoms in pixel. In our definition, all the atomic columns in the second layer (#2) should be visible thus fulfilling the desired pre-requisite to be fitted with Gaussian peaks for distance calculation. In other words, in the first atomic layer (#1) there are at least part of atomic columns not distinguishable in the ABF. To perform bond length calculation, we simply calculate the distance and convert the unit from pixel into pm (use the lattice constant of STO bulk as the reference to do the converting). Briefly explaining, to attain proper atomic position we have used Microsoft excel in order to sort the as-obtained random data first by x and then by y. Bond length is actually the distance between two adjacent atomic positions, for instance *m* and *m+1*, and we calculated it by using basic distance formula d = $\sqrt{(x_2 - x_1)^2 + (y_2 - y_1)^2}$ where $x_1$, $y_1$ and $x_2$, $y_2$ are the coordinates of *mth* and *(m+1)th* atomic positions respectively. Subsequently, similar procedure is adopted for all rows (meanwhile see Fig.1 for more clarity) consisting Sr-O and Ti-O plane along [001] direction towards surface. In addition to this, we repeat afore-mentioned process to obtain in-plane bond length for all columns including both SrO and TiO plane along [010] direction so that we could get lattice a which is an essential parameter to compute volume of unit cell for polarization calculation. Lattice c is obtained my merely adding two adjacent bond lengths along [001] direction than advanced on whole row and this procedure is gently applied on all layers. Bond length data was further used to find relative atomic displacement of atoms by using numerical differentiation method. For our calculations, $d_i$ *(i=1,2,3,4......20)* is the out of plane bond length where *i* is the sequence wise spacing among the layers ranging from one to twenty (also see Fig. 1(a)) and to calculate atomic displacement along [001] direction we use $\frac{|d_{i+1} - d_i|}{2}$ (where $d_i$ and $d_{i+1}$ are the adjacent bond lengths along c axis). Then we applied the same formulism for all the rows to get atomic shift as we did in case of bond length. Eventually, we use this obtained data of displacement along with magnitude of electronic charge (*e*), volume of unit cell ($a^2c$) and born effective charge values for Sr and Ti in order to calculate polarization by using the following precise relation $P_s = \frac{1}{V} \sum \delta_i Z_i$ [2], where V is the volume of unit cell for our case it is $a^2c$, $\delta$ is displacement/shift of atom (*i*) from their centrosymmetric position and Z is the Born effective charge of atom (*i*) calculated by ab initio theory having numerical value 7.12 for Ti and 2.54 for Sr [3]. Notably, small displacements (a few picometers) can exist in the bulk far away from the surface. These small systematic-error-displacements and polarization originate from the unavoidable specimen mistilt (a few mrad) between optical axis and specimen [4]. In order to extract the 'true' atomic structure, we use these systematic-error-displacements as a reference to calibrate the measured polarization, i.e., the mean of displacements far away from the surface is set to be zero as the base line. The values of Born effective charges for STO were previously calculated from ab initio theory [3].

### 3. Calculations among Ti-L edges from EELS data

The calculations are performed using the Digital Micrograph (DM, version 2.32.888.0, Gatan Inc.) software. To quantitatively obtained the intensity of Ti signals, first we calibrate the energy axis by the positions of Ti-L edges (456 eV) and O-K edge (532 eV) followed by the extraction of the Ti signal from the raw EEL spectra (use window of 405 ~ 455 eV). Secondly, we integrate the counts of Ti-L edges signal by the window of 456 ~ 470 eV.

For valence calculation, a build-in multiple linear least-squares (MLLS) fitting method of Digital Micrograph is used. In the preference of MLLS Fitting, we use 'Computer from Data' as the parameter of fit-weights and 'Fit coefficient' as the parameter of output fit. All the additional outputs are ticked on which are used to judge the accuracy of calculation. Next, we perform MLLS Fitting method on all the acquired EEL spectra (use window of 456 ~ 470 eV) with two standard EEL spectra of $Ti^{4+}$ and $Ti^{3+}$, which are normalized in advance. Using the result fit-coefficients of each spectrum, we can obtain the relative proportion of $Ti^{4+}$ and $Ti^{3+}$. Assuming the relative proportion of $Ti^{4+}$ and $Ti^{3+}$ as $a$ and $b$ ($0 \leq a, b \leq 1$), thus we can calculate the valence of $Ti^{x+}$ ($3 \leq x \leq 4$) of each experiment spectrum by the formula

$x = 4a + 3b$.

4. **Bond length values for top few unit cells**

Table 1: Average interlayer spacing down to three-unit cell length mapped along [001] direction. Bulk value of bond length for $SrTiO_3$, $d_o$=195.2 pm is used.

| Inter-layer Spacing | $d_{12}$ | $d_{23}$ | $d_{34}$ | $d_{45}$ | $d_{56}$ | $d_{67}$ |
|---|---|---|---|---|---|---|
| Magnitude(pm) | 189.2 | 214.6 | 209.5 | 210.5 | 186.8 | 201.0 |

## 5. Bond length measurement

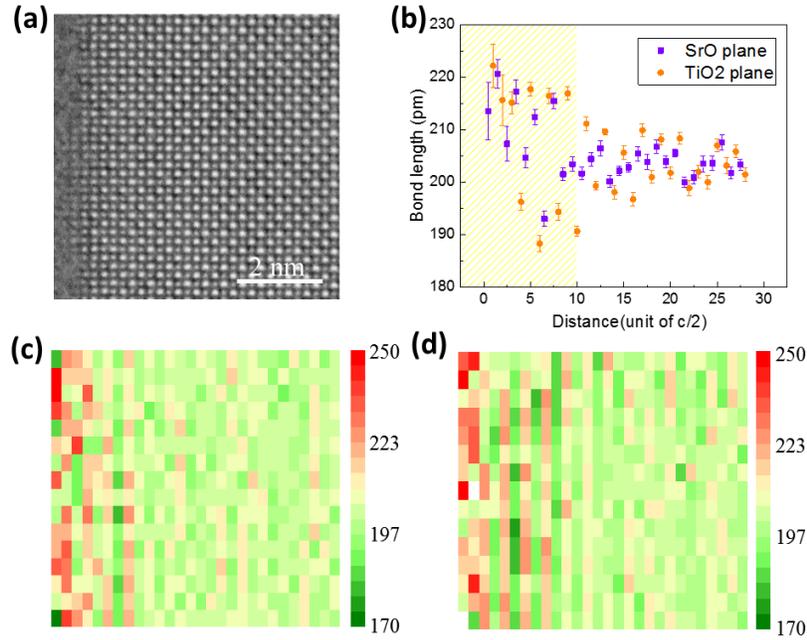

**Fig. S1** Illustration of atomic structure of SrTiO$_3$ surface and quantitative measurements of bond length for surface region of STO. The area of calculated region includes 16×13 number of unit cells. (a) An atomically resolved annular bright-field (ABF) image of SrTiO$_3$ (STO) surface. The atomic layers are alternative SrO and TiO$_2$ planes having viewing direction along [010]. (b) Mean of bond lengths. Twenty-seven data points where averaged for each plane to calculate bond length. The error bar is the s.d. Colored magnitude-map of bond lengths calculated at (c) SrO plane and (d) TiO$_2$ plane. The left and right side of these maps indicates surface and bulk region respectively.

## 6. Trend of lattice constant c

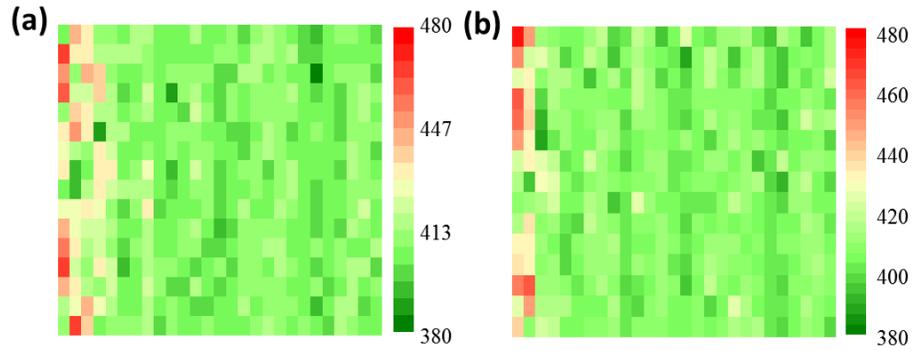

**Fig. S2** The magnitude map of lattice constant c at both Sr-O and Ti-O plane for an ABF image of SrTiO$_3$ (STO) surface shown in Figure S1. The magnitude map of lattice constant c calculated for (a) SrO plane, (b) TiO$_2$ plane. The left and right sides of these maps indicate surface and bulk regions respectively.

# 7. Quantitative measurements of structural parameters for another surface region

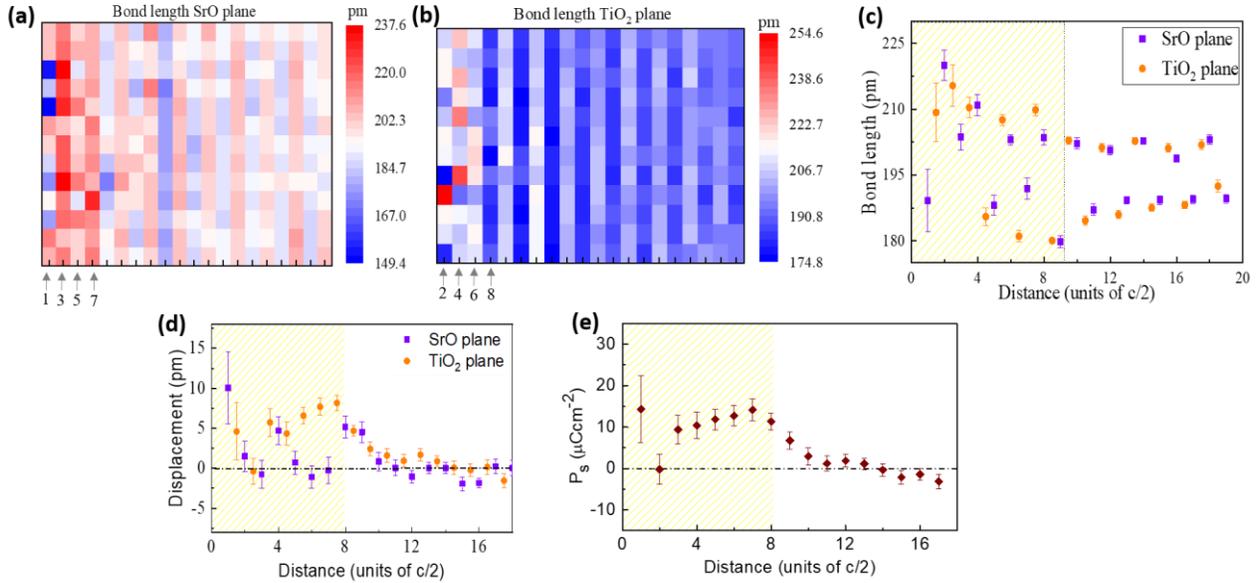

**Fig. S3** Quantitative measurements of structural parameters such as bond length, atomic displacement and polarization of surface region in STO thin film for an ABF image shown in Figure 1a. The area of calculated region includes 12×10 number of unit cells. Colored magnitude-map of bond lengths calculated at (a) SrO plane and (b) $TiO_2$ plane. The left and right side of these maps indicates surface and bulk region respectively. (c) Mean of bond lengths. Eighteen data points where averaged for each plane to calculate bond length. The error bar is the s.d. (d) Plot of z-component of atomic displacement at both SrO and $TiO_2$ planes along c axis. The error bar is the s.d. (e) Plot of total spontaneous polarization obtained from the displacement data mapped at SrO plane, $TiO_2$ plane. The error bar is the s.d.

## 8. Trend of lattice elongation for another surface region

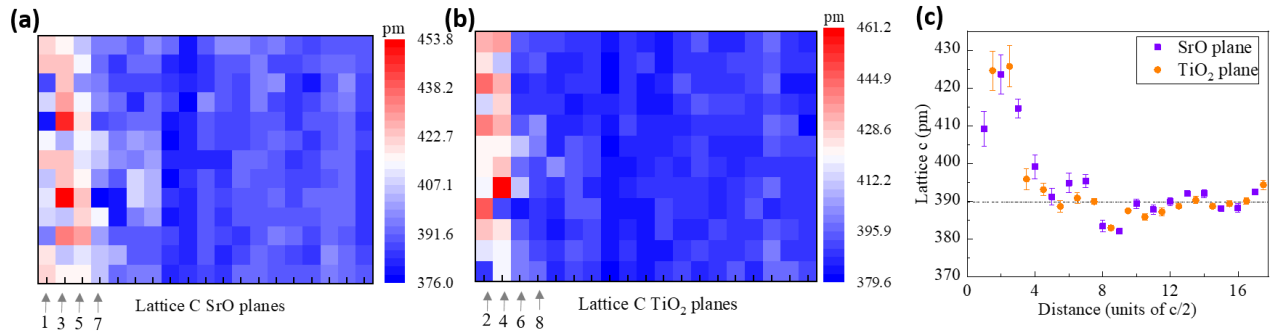

**Fig. S4** Trend of lattice elongation along crystallographic orientation c for under investigation STO thin film (image shown in figure 1). Colored magnitude-map of lattice c calculated at (a) SrO plane and (b) TiO$_2$ plane. (c) Plot for mean of the lattice parameter in [001] direction. The error bar is the s.d. Significant expansion of lattice c is being witnessed for skin as compared to bulk.

## 9. In plane, out of plane lattice constant and tetragonality measured for STO surface region

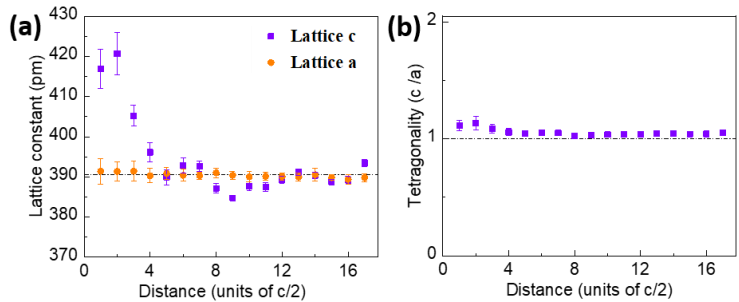

**Fig. S5** Plot for mean of the lattice parameter both out of plane and in plane. Tetragonality (c/a) is illustrated in (c).